\documentclass[10pt]{article}
\usepackage[OE]{express}

\begin{document}

\title{Demonstration of an efficient, photonic-based astronomical spectrograph on an 8-m telescope}

\author{N. Jovanovic,\authormark{1,2,*} N. Cvetojevic,\authormark{3} B. Norris,\authormark{4} C. Betters,\authormark{4}   \\ C. Schwab,\authormark{2,5}, J. Lozi,\authormark{1} O. Guyon,\authormark{1,6,7,8} S. Gross,\authormark{1,9} \\ F. Martinache,\authormark{10} P. Tuthill,\authormark{4}  D. Doughty,\authormark{1} Y. Minowa,\authormark{1} \\ N. Takato\authormark{1} and J. Lawrence\authormark{2,5}}

\address{\authormark{1}Subaru Telescope, National Astronomical Observatory of Japan, National Institutes of Natural Sciences (NINS), 650 North A'Ohoku Place, Hilo, HI, 96720, U.S.A.\\
\authormark{2}MQ Photonics Research Centre, Department of Physics and Astronomy, Macquarie University, NSW 2109, Australia\\
\authormark{3}Observatoire de Paris, Meudon, 5 Place Jules Janssen, 92195, France\\
\authormark{4}Sydney Institute for Astronomy (SIfA), Institute for Photonics and Optical Science (IPOS), Sydney Astrophotoinc Instrumentation Laboratory, School of Physics, University of Sydney, NSW 2006, Australia\\
\authormark{5}Australian Astronomical Observatory, 105 Delhi Rd, North Ryde NSW 2113, Australia\\
\authormark{6}Steward Observatory, University of Arizona, Tucson, AZ, 85721, U.S.A.\\
\authormark{7}College of Optical Sciences, University of Arizona, Tucson, AZ 85721, U.S.A.\\
\authormark{8}Astrobiology Center of NINS, 2-21-1, Osawa, Mitaka, Tokyo, 181-8588, Japan\\
\authormark{9}Centre for Ultrahigh-bandwidth Devices for Optical Systems (CUDOS)\\
\authormark{10}Laboratoire Lagrange, Universit\'{e} C\^{o}te d'Azur , Observatoire de la C\^{o}te d'Azur, CNRS, Parc Valrose, B\^{a}t. H. FIZEAU, 06108 Nice, France}

\email{\authormark{*}jovanovic.nem@gmail.com} 


\begin{abstract}
We demonstrate for the first time an efficient, photonic-based astronomical spectrograph on the 8-m Subaru Telescope. An extreme adaptive optics system is combined with pupil apodiziation optics to efficiently inject light directly into a single-mode fiber, which feeds a compact cross-dispersed spectrograph based on array waveguide grating technology. The instrument currently offers a throughput of $5\%$ from sky-to-detector which we outline could easily be upgraded to $\sim13\%$ (assuming a coupling efficiency of $50\%$). The isolated spectrograph throughput from the single-mode fiber to detector was $42\%$ at $1550$~nm. The coupling efficiency into the single-mode fiber was limited by the achievable Strehl ratio on a given night. A coupling efficiency of $47\%$ has been achieved with $\sim60\%$ Strehl ratio on-sky to date. Improvements to the adaptive optics system will enable $90\%$ Strehl ratio and a coupling of up to $67\%$ eventually. This work demonstrates that the unique combination of advanced technologies enables the realization of a compact and highly efficient spectrograph, setting a precedent for future instrument design on very-large and extremely-large telescopes.  
\end{abstract}



\section{Introduction}
The field of astronomical spectroscopy is rapidly approaching an impasse: the size and cost of instruments for extremely large telescopes (ELTs) is pushing the limits of what is feasible as they require optical components that are at the very edge of the physical size that can be achieved. For this reason astronomers are constantly looking to expand their arsenal by embracing new technologies. This includes photonic technologies that provide an avenue to miniaturization and simplification. Although astronomers have long been curious of the potential of photonics~\cite{coude94}, there was a renaissance around $2009$ where the new sub-field of "astrophotonics" was coined, which has been embraced by the community~\cite{jbh2009}. 

Photonic technologies can provide advanced functionalities like spectral filtering~\cite{jbh2011,mar2012,spal2014}, spectral dispersion~\cite{cvet2009}, frequency comb generation~\cite{feger2014b,schwab15} and spatial filtering~\cite{coude94}, and do so by operating at the diffraction limit. Besides offering these functionalities, operating at the diffraction-limit can overcome classical instrument scaling laws (carefully outlined in~\cite{Jov2016}). The prospect of miniaturization of astronomical instrumentation has been one of the primary motivators thus far. Another key motivator is provided by spatial filtering. This property ensures that the output beam from a single-mode fiber (SMF) is temporally invariant in regards to its shape (the amplitude can vary), which means the point-spread function (PSF) inside the instrument is well understood and can be easily calibrated out. This is an extremely powerful property that would eliminate modal noise in spectrographs, for example~\cite{schwab2012}. For both of these two reasons, several groups have recently deployed diffraction limited spectrographs on $8$-m class telescopes~\cite{rains2016,crepp16}.        

However, coupling into a SMF from a large telescope from the ground has historically been very difficult, due to atmospheric turbulence and the mismatch between the telescope beam and the profile of the fundamental mode of a SMF~\cite{Coude2000,woillez2003}. Second generation adaptive optics systems have been instrumental in improving coupling efficiencies for both interferometery~\cite{menn2010} and most recently spectroscopy~\cite{Bechter2016}, where a coupling efficiency of $20$--$25\%$ in y-band was demonstrated. In an attempt to address this short coming and open the possibility of exploiting photonics efficiently on larger telescopes, we investigated the possibility of using advanced wavefront control, combined with lossless pupil apodization. Recently, we successfully demonstrated that by using these technologies, it was indeed possible to get very high levels of coupling ($>50\%$) into a SMF from an $8$-m telescope~\cite{jov2017}. This demonstration makes it possible to conceive of efficient photonic/diffraction-limited instrumentation for larger telescope apertures, for scientific use, for the first time. Indeed, this formed the motivation for the work we present here.

The spectrograph was based on an arrayed waveguide grating (AWG) chip, which we had previously characterized in the laboratory~\cite{cvet2012a}. That device was tested in an early prototype instrument on the Australian Astronomical Telescope (AAT)~\cite{cvet2012b}. The $3.9$~m telescope is seeing-limited and thus the instrument required a photonic lantern~\cite{saval2013} to couple the light to the chip efficiently. This instrument architecture was first proposed and described in~\cite{jbh2010} and dubbed the PIMMS concept. Despite the fact that spectra were successfully collected, the prototype instrument was unoptimized and hence inefficient and simply not viable for science operation. However, now that we have recently demonstrated efficient coupling to SMFs on-sky, our aim was to take the concept one step further and feed the AWG spectrograph with a single SMF. Here we present the first highly efficient photonic-based spectrograph, which can be fed directly from a large telescope. Section~\ref{experiments} outlines the experimental setup and instrument overall. Section~\ref{sec:results} summarizes the key results including the throughput, resolving power and offers some stellar spectra while Section~\ref{sec:conclusion} rounds out the paper with some concluding remarks.

\section{Experimental setup}\label{experiments}
The experiments were conducted with the aid of the Subaru Coronagraphic Extreme Adaptive Optics (SCExAO) instrument, which is based at the 8-m Subaru Telescope, on Maunakea, Hawaii. The SCExAO instrument provided the high level of wavefront correction required to efficiently inject light into a SMF. The light was dispersed by a simple spectrograph based on an AWG device in the near-IR (NIR). This section outlines the various aspects of the setup including the spectrograph. 

\subsection{The adaptive optics system}
\begin{figure}[!t]
\centering 
\includegraphics[width=0.99\linewidth]{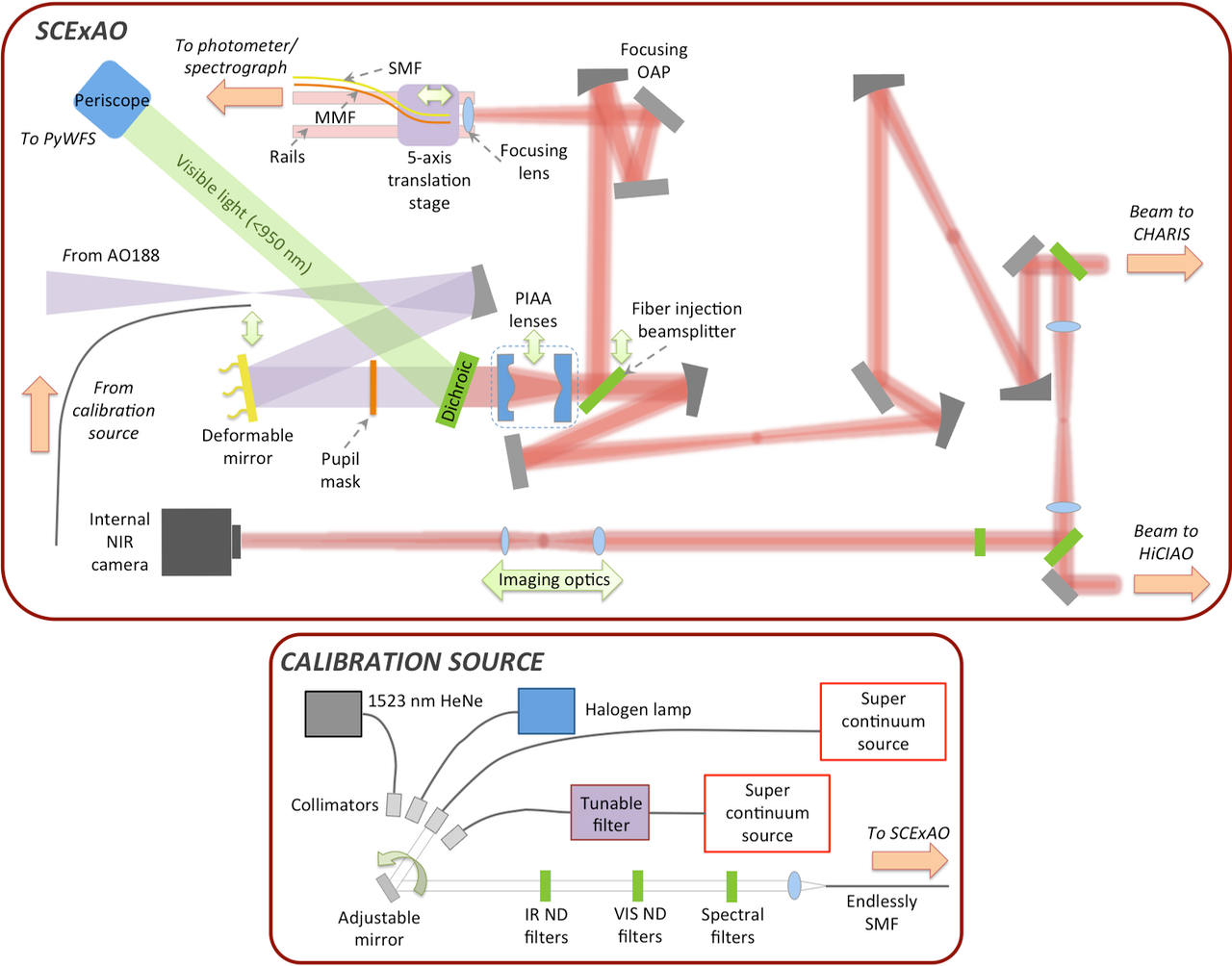}
\caption{\footnotesize (Top) Schematic diagram of the IR bench of the SCExAO instrument. (Bottom) Schematic diagram of the portable calibration source. Dual head green arrows indicate that a given optic can be translated in/out of or along the beam. Orange arrows indicate light entering or leaving the designated bench at that location. IR-infrared, VIS-visible, ND-Neutral density.}
\label{fig:schematic}
\end{figure}

To achieve efficient coupling into a SMF, it is imperative to flatten the wavefront in order to match that required by the fundamental mode of a SMF (i.e. a flat phase front). The SCExAO instrument was utilized to achieve this by virtue of its advanced wavefront control. Here we focus only on the aspects that are relevant to this body of work and direct the reader to the literature for further information about the instrument~\cite{jov2015}. The light enters SCExAO from AO188 (as shown in Fig.~\ref{fig:schematic}) an upstream adaptive optics (AO) system that provides an initial correction to the star light: this system achieves typical Strehl ratios of $30$--$40\%$ in the H-band ($1490$--$1800$~nm) in median seeing (0.65 arcsecs) conditions~\cite{min2010}. The Strehl ratio is a measure of image quality and can be calculated as the ratio of the peak flux in an image to the peak flux for a perfect image (i.e. free from wavefront error). With this definition a perfect image would have a Strehl ratio of $1$ (or $100\%$) and decrease as the amount of wavefront error increased to a minimum of $0$. The light is next sensed by the pyramid wavefront sensor (PyWFS), which operates from $800$--$900$~nm. The PyWFS drives the $2000$ actuator deformable mirror (DM) inside SCExAO to further improve upon the correction provided by AO188. It does this by sensing and correcting for both higher spatial and temporal frequencies of the atmospheric turbulence profile than AO188 can. It regularly operates at a loop speed of $2$~kHz, with a latency of $\sim1$~ms on up to $1300$ modes. It is still undergoing commissioning but has achieved up to $80\%$ Strehl ratio in the H-band in very good conditions (better than median seeing). More typically however, it provides between $60$--$70\%$ Strehl ratio performance. For completeness the PyWFS has a tip/tilt mirror in an upstream pupil plane for modulation. Note, none of the hardware for the PyWFS are shown in Fig.~\ref{fig:schematic} as it is mounted on a second bench. 

\subsection{The fiber injection unit}
Within the SCExAO instrument, a fiber injection unit was built to couple the light into a SMF. This unit is described in detail in~\cite{jov2017}, and here we simply highlight the key features. Prior to injecting the light into a fiber, it is important to convert the top-hat beam which produces an Airy pattern at the focus of the telescope to a Gaussian to better match the collection fiber. By doing this it is possible to eliminate the $60\%$ coupling limitation for the Subaru Telescope pupil geometry ($30\%$ central obstruction and thick spiders) and elevate it to $91\%$, in the diffraction limit. This was achieved by utilizing apodization optics, which consist of two aspheric lenses that remap the flat-top illumination of the pupil into a prolate spheroid profile (a near-Gaussian shape). The lenses are made from CaF$_{2}$ and have one flat and one aspheric surface each~\cite{lozi2009}. The first lens begins the process of apodizing the pupil by pushing rays towards the center of the pupil while the second lens simply recollimates the beam. This generates a quasi-Gaussian beam in the focus that is a better match to the fiber mode. 

The fiber injection unit consists of a combination of an off-axis parabolic (OAP) mirror and a small aspheric lens, which can be used to modify the focal ratio of the beam by altering their relative spacing. The focal ratio was set to $\sim5.3$ in order to optimize coupling into the SMF-28-J9 fiber used in these experiments. The fiber was mounted on a 5-axis stage (Newport, M-562-XYZ and 562F-TILT), which offered the ability to tune the X, Y and focus axes of the fiber in the focal plane of the lens system via computer controlled stepper motors (Zaber, T-NA08A25). The actuators offered a minimum step size of $50$~nm and an unidirectional repeatability of $<1~\mu$m. Tip/tilt utilized manual actuation and was pre-aligned in the laboratory by hand. A large multimode fiber (MMF, $365~\mu$m core diameter, $0.22$ NA, step index) was co-mounted on the same translation stage and was used to calibrate the absolute coupling into the SMF. This was done by simply translating from the SMF to the MMF on a given target and taking the ratio of the two signals (i.e. the assumption was that the MMF collected all of the light). All coupling efficiencies measured using this method were reported in~\cite{jov2017}, here we simply reuse the values.   

The fiber injection was thoroughly characterized and the results were presented in~\cite{jov2017}. Here we provide an overview. The fiber injection unit achieved a peak coupling efficiency of $78\%$ between $1500$--$1600$~nm in the laboratory (i.e. with a perfectly flat wavefront). This value dropped by $\sim30\%$ at shorter wavelengths ($1250$~nm). The coupling efficiency ($\eta$), which is defined as the ratio of the power inside the fiber with respect to that incident on the fiber (removing the effect of Fresnel reflection from the front surface of the fiber) was linearly correlated with the Strehl ratio in that work and the equation of the fit is given by $\eta = Strehl~Ratio \times0.74+1.84$ ($\%$). Note the Strehl ratio needs to be input as a percentage and $\eta$ will be expressed as one. Preliminary on-sky testing with a Strehl ratio of $\sim60\%$ in the H-band yielded a measured average coupling efficiency of $47\%$. This value was consistent with expectation from the linear correlation validating it. This means that the coupling efficiency is well understood and the performance on a given night is solely determined by the Strehl ratio of the beam used for injection. This is an ideal scenario as it offers us the possibility to forecast the system performance given the correction provided.  

\subsection{The spectrograph}
\begin{figure}[!b]
\centering 
\includegraphics[width=0.99\linewidth]{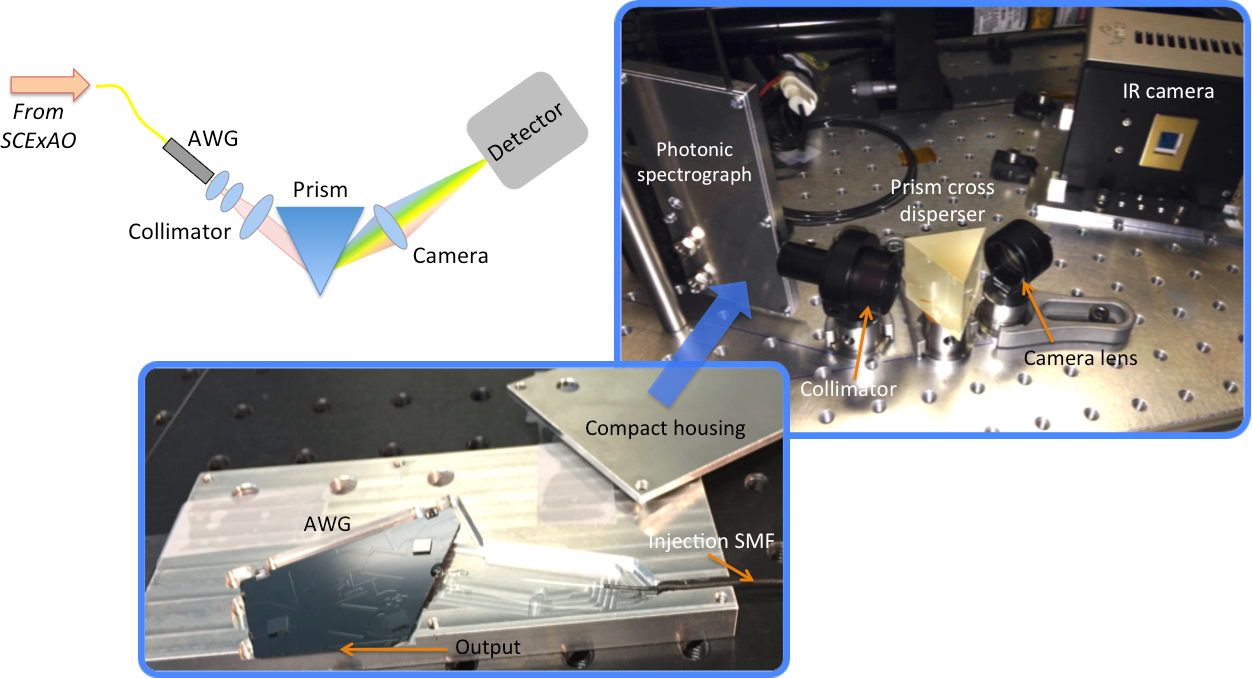}
\caption{\footnotesize (Top Left) Schematic of the photonic spectrograph. (Top Right) A laboratory image of the photonic spectrograph highlighting its simplicity. (Bottom) The AWG mounted inside a compact housing with a single SMF for delivering the light.}
\label{fig:spectrograph}
\end{figure}

The light from the SMF was split into a $90/10$ ratio with an achromatic fiber splitter (TW1550R2F2, Thorlabs). The $10\%$-port was routed to a photometer used to measure the coupling (discussed in~\cite{jov2017}). The $90\%$-port was fed to a compact spectrograph, shown in Fig.~\ref{fig:spectrograph}. 
The fiber was directly bonded to an AWG spectrograph chip with UV curing epoxy. This is the same device presented and thoroughly characterized in~\cite{cvet2012a}. An AWG achieves spectral dispersion by passing the light through an array of single-mode waveguides constructed such that each waveguide is a fixed path length longer than its nearest neighbor. In this way the waveguide array is analogous to the teeth of a blazed grating, and diffracts light at different wavelengths into different directions at the output of the array. For a more detailed description we refer the reader to~\cite{cvet2012a,lawrence2010}. The key properties of the device include a resolving power of $R=7000$, and a free spectral range (FSR) of $52$~nm at the central wavelength of operation of $1550$~nm. In order to increase the spectral coverage of the device beyond a single FSR, the output beam was cross-dispersed onto a detector. The light was collimated by three, $f=50$~mm achromats, before being dispersed by an equilateral NSF11 prism (uncoated). Three doublets were needed to minimize the chromatic aberrations across the field. The output beam was focused with a $f=150$~mm achromat onto an InGaAs detector (Xeva-1:7-320 TE3, Xenics). The $320\times256$ pixel camera had $30~\mu$m square pixels and employed a $3$-stage thermoelectric cooler (TE$3$). It had $\sim170$~electron read noise and $\sim3\times10^{4}$~electrons/pixel/second dark current, which limited the integration time $\sim6$~s before the well was full, in high gain mode. This detector does not have the performance of science grade detectors like HAWAII $2$RGs but was accessible for these experiments. The detector was inclined at $45^{\circ}$ with respect to the beam for optimum focus of the spectra at all wavelengths simultaneously, given the large chromatic defocus still remaining in the optical system.

To maximize stability, the beam height was set to $50$~mm from the breadboard, with fixed height posts/spacers and the breadboard was reinforced with $50$~mm wide struts from beneath. The optics were enclosed with a small enclosure that consisted of $5$~mm thick cardboard panels. The cooling fan of the IR camera was placed on the outside of the enclosure to prevent heat being dumped inside the instrument box. No measures were taken to thermally stabilize the spectrograph. It was located at the Nasmyth platform of Subaru Telescope for the duration of these experiments. This environment can oscillate in temperature by $5^{\circ}$C over the course of 24 hrs depending on whether the air conditioning is on or not. This manifested in drifts of the spectrograph that shifted the traces by up to several pixels per night.

\section{Results}\label{sec:results}
\subsection{Overall system efficiency}\label{through}
In order to justify the use of a diffraction-limited spectrograph for astronomy it is important that the total system efficiency, from sky to the detector focal plane be relatively high and ideally similar to a typical MMF-fed spectrograph. In this vein, the throughput of the individual elements between the sky and the plane of the detector are summarized in table~\ref{tab:throughputs}. 

All throughputs for the pre-injection elements including the sky, telescope, ADC, AO188, SCExAO optics and apodization optics were taken from~\cite{jov2015}. These indicate that only $25\%$ of the flux from the sky makes it to the focal plane of the fiber. The largest contribution to loss from these elements comes from the apodization optics. The reason for this is that we currently use a lossy binary mask to aid the lenses in the apodization process~\cite{lozi2009}. It is worth noting that the coupling improvement gained by using the current optics (i.e. increasing the limit from $60\%$ to $91\%$) is outweighed by the losses in the optics themselves, so there is no advantage of using them currently. New aspheric optics have recently been designed that don't rely on the mask (please see~\cite{jov2017} for further details). These will boost the throughput by almost a factor of $2$, increasing the overall efficiency of the pre-injection to $45\%$. With the new optics there will be a benefit to utilizing the optics. Nonetheless, here we demonstrate the entire optical train as we envision it with the current optics to demonstrate the feasibility of the technology.  

\begin{table}[ht!]
\centering
\caption{Throughput of the elements of the photonic spectrograph at $1550$~nm. S.R. - Strehl ratio, ADC - Atmospheric dispersion compensator. Values for the throughput of the atmosphere, telescope, ADC, AO188, SCExAO and apodization optics were taken from~\cite{jov2015}. The coupling was taken from~\cite{jov2017}. The AWG throughput was taken from~\cite{cvet2012a}. The total throughput assumes a coupling efficiency of $50\%$ (corresponding to $65\%$ Strehl ratio at H-band).}
\begin{tabular}{|l|c|c|}
\hline
\textbf{Element}		& \textbf{Throughput (\%)}  	& \textbf{Optimized throughput (\%)} \\ \hline
\textbf{Pre-injection}  & 	 							& 	 \\
Atmosphere 	            & 97 							& 97 \\
Telescope 	            & 92 							& 92 \\
ADC						& 92 							& 92 \\
AO188		            & 79 							& 79 \\
SCExAO		            & 68 							& 72 \\
Apodization optics		& 55 							& 96 \\ \hline
\textbf{Throughput to injection} & \textbf{24}			& \textbf{45} \\ \hline
\textbf{Post-injection}	&								&	 \\
Transmission at fiber interface & 96							& 0.995 \\ 
AWG 		  			& $77\pm5$						& $77\pm5$ \\ 
AWG bonding/splices		& 79							& 95 \\
Collimating lens     	& 90 							& 90 \\ 
Prism			        & 82 							& $>$90 \\ 
Camera lens	            & 97 							& 97 \\ \hline
\textbf{Spectrograph throughput} & \textbf{42}			& \textbf{57} \\ \hline
Coupling				& S.R.$\times 0.74+1.84$  		& S.R.$\times 0.74+1.84$ \\ \hline
\textbf{Total throughput} & \textbf{5}					& \textbf{13} \\
\hline
\end{tabular}\label{tab:throughputs}
\end{table}

The throughput of the post-injection elements was measured by injecting light from a $25$~nm band of the super-continuum laser (Fianium-Whitelase micro), centered on $1550$~nm, into the input SMF and measuring the flux after each optical element in turn. The spectrograph itself showed an excellent throughput of $42\%$. Despite this there are several key areas which can be improved to maximize the overall throughput. Firstly, the collection fiber could be AR-coated. Secondly, the alignment of the SMF with the AWG, subsequent bonding to the chip and splicing of the SMF to a connectorized fiber currently account for $21\%$ loss. This value could be reduced to $\sim5\%$ with a more careful alignment of the fiber to the AWG, and a higher quality splice (a lossless splice is possible in this type of fiber). Finally, the NSF11 prism did not have an AR coating. As the beam was at a high angle of incidence ($\sim60^{\circ}$) with respect to the normal of the prism, the Fresnel reflection losses were high and polarization dependent. The value reported in the table, $\sim82\%$ is for the unpolarized beam of the super continuum source (i.e. average across both polarizations). Implementing AR coatings can improve this value. But at the large angle of incidence it is hard to predict the exact performance so we place a lower limit of $90\%$ throughput for such an optic with AR coatings. With relatively simple upgrades the throughput of the spectrograph can be increased to $57\%$ in future. Nonetheless the system that we tested was certainly very efficient already.    

If we assume a conservative Strehl ratio of $65\%$ in H-band can be achieved, which takes into account sub-optimal conditions, then the current total system efficiency from sky-to-detector is $5\%$. With the upgrades described above it could be up to $13\%$. These values are commensurate with typical MMF fed spectrographs. For example, the precision radial velocity machine IRD is forecast to have a sky-to-detector throughput of $6\%$\cite{tamura2012}, which means that this diffraction-limited instrument would be a competitive alternative in some areas of astronomy. These areas are limited to targets brighter than $9^{th}$ magnitude in the I-band ($730$--$880$~nm), which is a requirement for the AO system to achieve the level of Strehl ratio assumed here. This limiting magnitude would not be improved on if designing a similar system for an ELT. This seems counter intuitive at first but the rate at which the signal would increase due to the larger aperture would be canceled out by the fact there would be more wavefront (across the aperture) to control and hence the signal per sub-aperture would be constant~\cite{guyon2005}. In addition, an instrument like this would be limited to point sources, and so could not be used to study an entire galaxy at once. It could however, be used to scan across extended sources such as galaxies. Given these constraints, the instrument presented here could be used to deliver medium-resolution spectra of imagable exoplanets currently studied with integral field spectrographs~\cite{barman2015}. Several other applications become possible if the AWG were to be replaced with one with a greater resolving power (R>50000), currently under investigation~\cite{cvet2012c}. For example, this instrument architecture is best suited to precision radial velocity surveys that require the highest precisions as outlined in~\cite{Crepp2014,Jov2016}. Another ideal application is that of high dispersion coronagraphy, whereby the light from a known exoplanet is collected by a fiber from a post-coronagraph focal plane where the photon noise has been greatly suppressed. This emerging application is a promising approach to obtain high-resolution spectra of imagable exoplanets and is enhanced by utilizing a SMF to feed the spectrograph~\cite{Wang2017, Mawet2017}. 

Although we have chosen to itemize the losses and group them in regards to pre and post-injection elements, the instrument really begins at the input to the first AO system and finishes at the detector. Both AO systems are required to achieve the high coupling efficiency needed to exploit the photonic spectrograph so it is instrumental to think about the system as a whole. The throughput of the entire instrument is $6\%$ with the current optics and could be elevated to $16\%$ with the upgrades outlined in future. 

The values reported above exclude the quantum efficiency of the detector. As outlined we used a research grade detector and instruments optimized for science would typically rely on a lower noise detector like a HAWAII 2RG array. In addition, the throughputs presented above are at a single monochromatic wavelength. One must consider the efficiency across the entire band of operation when designing an instrument. We use the data at the single wavelength as a yard stick for the system throughput of this combination of technologies. There is no fundamental reason why a similar level of performance could not be achieved across a broad range of wavelengths.

\subsection{Spectrograph properties}
\begin{figure}[!b]
\centering 
\includegraphics[width=0.99\linewidth]{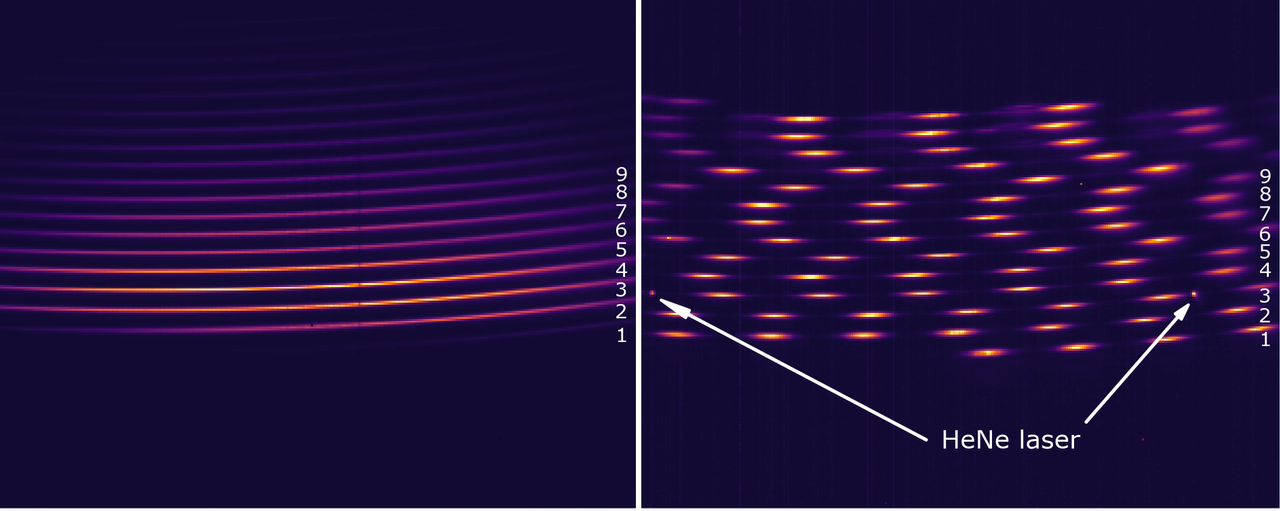}
\caption{\footnotesize (Left) An image of the super continuum source illuminating the spectrograph. (Right) A co-added image of most of the spots created by scanning a narrowband laser from $1110$ to $1670$~nm. The image of the $1523$~nm HeNe laser is superimposed. The short wavelength end of each order is to the right of the image, while the shorter wavelength orders are towards the top of the images. The orders that were extracted are numbered to the right of each image.}
\label{fig:labspec}
\end{figure}

Before using the spectrograph it is important to first characterize the basic properties of the instrument including the wavelength range and the resolving power. Initially, the spectrograph was illuminated with the broadband spectrum of a super continuum source. An image depicting the multiple orders of the spectrograph on the detector is shown in Fig.~\ref{fig:labspec}. Each trace in the image corresponds to a different order.   
To determine the location of each wavelength and the resolving power, a narrowband light source was used. This was created by passing the light of a second super-continuum (Fianium - Whitelase SC480) source through a tunable filter (Fianium - LLTF SWIR, bandwidth of $\sim5$~nm). The filter was adjusted in steps of $10$~nm from $1110$~nm up to $1670$~nm and images were acquired. A combined image with most of the spots during the wavelength scan is shown in the right panel of Fig.~\ref{fig:labspec}, which includes the spots from the $1523$~nm HeNe laser superimposed for reference. It should be made clear that the integration time on the detector was adjusted as the wavelength was scanned to keep the signal at a near constant level throughout the experiment. This explains the relative gradient in brightness across the orders between the left and right panels. It is immediately obvious from the right panel that the spectrograph can easily detect the difference between the $5$~nm linewidth of the tunable laser and that of the much narrower HeNe laser. A thorough measurement of the resolving power/resolution are presented below. It is also clear that the $1523$~nm HeNe laser was on the edge of a FSR and can be seen in two neighboring orders, which aided in absolute wavelength referencing of these two orders.    

Data extraction was implemented in a multi-step process utilizing the pipeline first developed in~\cite{betters2013}, which is based on optimal extraction~\cite{horne1986}. Some modifications to that pipeline were implemented for this work/instrument. The pipeline was not optimized as would be the case for a science ready instrument and is an area for future upgrades. However, it served the purpose of demonstrating the functionality and performance of the instrument in this work. 

The data extraction/reduction process is outlined as follows. Firstly, dark frames taken at the same exposure time were collapsed into a single average dark frame, which was subtracted from all data frames. Hot pixels were removed and the images were destriped. Destriping is an important step to remove residual vertical stripes common to NIR detectors, and if neglected will result in high frequency modulation along the spectrum. 

The second step involved extracting each trace. Multiple Gaussian profiles were fitted to the traces along a column of the detector (orthogonal to the spectral direction). This was repeated across the entire array (each column). The center of each Gaussian along a single trace was fitted by a multi-order polynomial to map the location of that trace. The fitted height of each Gaussian was extracted (as the flux) and plotted as a function of the pixel coordinate along the horizontal axis of the detector (column number). Note, to get the highest precision fit, all pixels along each trace had to be illuminated by either using the broadband super continuum source or the halogen lamp. Due to the faintness of some orders towards the top of the image in the left panel of Fig.~\ref{fig:labspec}, and the unoptimized pipeline, the extraction process was only used on the $9$ brightest orders which returned robust fits. The $9$ orders extracted throughout this work are labeled in the images in Fig.~\ref{fig:labspec}. Once the polynomial fit was determined, it was used to extract the traces for the wavelength reference data shown in the right panel of Fig.~\ref{fig:labspec} above. Note, this assumes there was no drift in the image on the detector, which is one area that could be improved in future. 

The third step involved flat fielding. Once a line profile was produced along a single trace, which had peaks corresponding to the location of each position of the tunable filter, it was divided by the response of the AWG order which was measured using the halogen lamp. This simply removed any low order variations across the trace. 

\begin{figure}[!t]
\centering 
\includegraphics[width=0.99\linewidth]{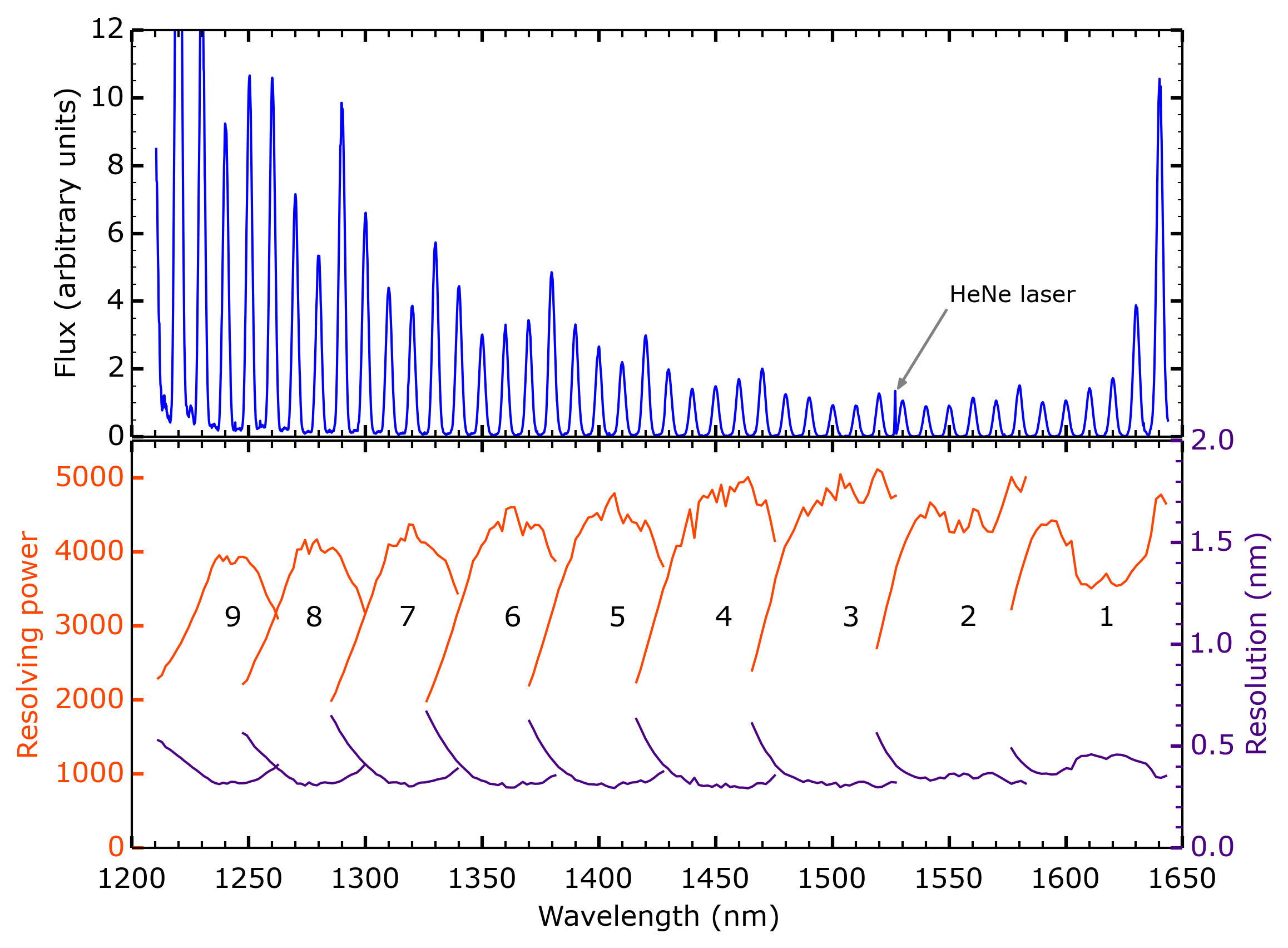}
\caption{\footnotesize The resolving power and resolution as a function of the wavelength for the $9$ extracted orders. The order number is shown in the figure.}
\label{fig:resolve}
\end{figure}

As a final step a wavelength solution for each peak, and each FSR was developed by using the knowledge of the wavelength of the tunable laser. This generated a single continuous spectrum that spanned from $1220$--$1650$~nm, shown in the top panel of Fig.~\ref{fig:resolve}.   

The width of each Gaussian fitted to the spectral traces was used as a proxy for the spot size in that location of the image. The rationale for doing this is that light is injected into the AWG using a SMF which delivers a circularly symmetric Gaussian profile. When the AWG chip is used in isolation, the beam at the input (from the SMF) is preserved for an on-axis launch~\cite{cvet2012a}. Hence, by measuring the vertical width of a trace, which we assume to be circularly symmetric, we gain insight into the relative spot size (FWHM) across the detector. The resolving power across the detector was determined by finding the ratio of the spot size (in pixels) at each location in each trace and normalizing it by the distance between the two closest consecutive wavelength scan spots (in pixels), multiplied by the wavelength step size used in the scan (in nm). The resolving power as a function of the wavelength is plotted in the bottom panel Fig.~\ref{fig:resolve}.

It can be seen that the peak resolving power in the $9$ orders that were extracted ranges between $4000$--$5000$. The native resolving power of the AWG ($7000$) is slightly reduced when reimaged onto the detector. In addition, the resolving power for each order is asymmetric. It can be seen that it is systematically lower to the shorter wavelength end of each FSR. This can be understood by looking at the right panel of Fig.~\ref{fig:labspec} above, where it can be seen that the spot on the right hand side of the image (i.e. shorter end of each FSR) are not as well focused as the spots on the left of the image. Hence, the focus of the spots across the array was not perfect. The corresponding resolution is also seen in the bottom panel of Fig.~\ref{fig:resolve}. It can be seen that it is between $0.3$--$0.4$~nm across most of the spectrum.

The extracted line profile along the composite image of all the traces shows that each peak is quasi-Gaussian and has a single peak. This indirectly validates that the AWG remained single-mode even at the shorter wavelengths. The SMF has an upper limit specification of the second-mode cutoff wavelength of $\sim1250$~nm. This means that depending on the batch, the fiber could support $2$ modes or more at any wavelength below this. It may have been possible to excite the second mode of the fiber at shorter wavelengths but for them to be rejected by the AWG (i.e. they may have higher loss). It is also entirely possible that because of the accurate attempt to match the spot size and shape, we may have only excited a single mode in the SMF and it remained that way throughout the instrument. Regardless, from our experiments, the system seems to operate on a single mode for wavelengths $>1220$~nm. The long wavelength end is limited by the near-zero quantum efficiency of the detector above $1650$~nm. It is unclear which component would impose the next limit to the longest wavelength possible given a detector with a broader range was used instead. 

Interestingly, although the tunable filter could not be set below $1100$~nm, light was seen on the detector down to $1000$~nm when a broadband source was used. This indicates that it might be possible to use this device at even shorter wavelengths, although the location of the single-mode cutoff of the AWG is not known and may impose a lower limit. This is large wavelength range for a photonic chip that was not optimized for this application. The device is simply an off-the-shelf item used in telecommunications most commonly.

\subsection{On-sky results}
In order to test and demonstrate the spectrograph, spectra of stars were collected on-sky. It is important to note that access to telescope time was scarce and only available during SCExAO commissioning/engineering nights. This meant that we could not select periods with ideal conditions for these tests. Therefore, the results presented here were taken in low to medium Strehl ratio conditions. The fiber injection has however been tested in both the laboratory and more relevantly on-sky with Strehl ratios of up to $60\%$ (in H-band) validating that the single greatest risk to this instrument concept, efficient injection into a SMF, does indeed work and is well understood (note the Strehl ratio in J-band was up to $15\%$ lower). Here we simply present spectra from a range of stellar targets to demonstrate the successful operation of the spectrograph.

\begin{figure}[!b]
\centering 
\includegraphics[width=0.99\linewidth]{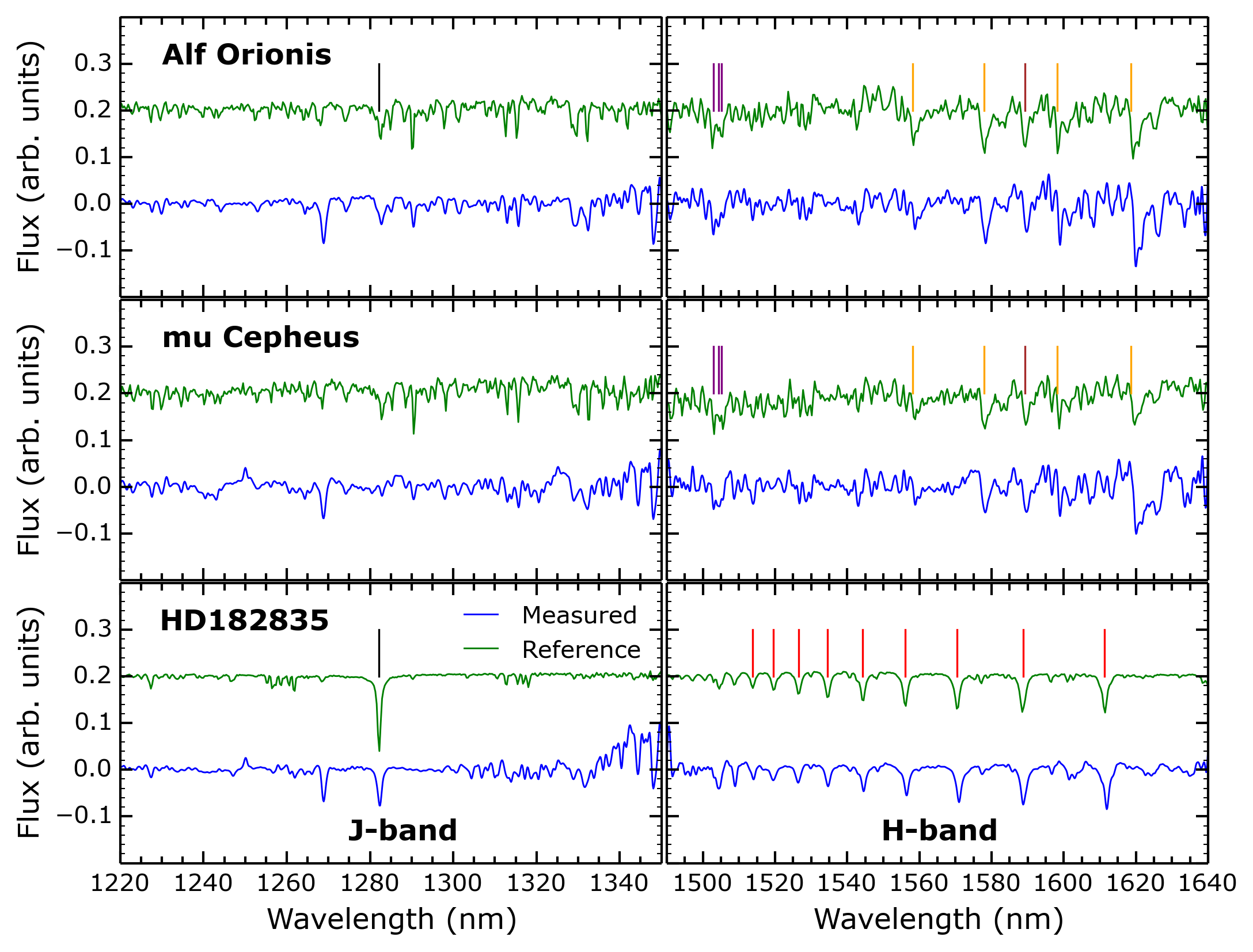}
\caption{\footnotesize (Left) J-band and (Right) H-band spectra of (Top) Alf Orionis (Middle) mu Cepheus and (Bottom) HD182835. The reference spectra were taken from~\cite{rayner2009}. Red vertical lines indicate the location of hydrogen-Brackett lines. Black vertical lines indicate the location of hydrogen-Paschen lines. Orange vertical lines indicate the location of CO lines. Brown vertical lines indicate the location of silicon lines. Purple vertical lines indicate the location of magnesium lines.}
\label{fig:onskyspec}
\end{figure}

The first data set was collected on the night of the $28^{th}$ of October, $2015$ on the star Alpha Orionis (spectral type M1, H magnitude of $-3.73$). The second was taken on July $29^{th}$, $2015$ on the star mu Cepheus (spectral type M2, H magnitude of $-1.27$). A third data set was collected on the same night as mu Cepheius on the star HD182835 (spectral type F2, H magnitude of $3.0$). The J and H band spectra are shown in the left and right panels of Fig.~\ref{fig:onskyspec}, respectively. 
The data is shown in blue while a reference spectra taken from the IRTF catalog are shown in green~\cite{rayner2009}. Note that the reference spectra have been normalized to remove the black body profile for the star. The measured spectra were convolved with a Gaussian to down sample them to the resolving power of the SpeX spectrograph used at IRTF (resolving power of $2000$) for better comparison. An additional low order term was used to remove systematics from the collected data. The need for this additional term is due to an unoptimized data extraction and reduction routine, which can easily be remedied in future.

It can be seen that the spectra appear to be qualitatively and quantitatively similar to the reference spectra, which is encouraging. Differences between the spectra could come from three distinct sources, which include, a difference in the resolving power of the two spectrographs used to collect them, the quality of the calibration procedure and/or fluctuations in the spectrum of the star. As outlined the data was convolved with a Gaussian, which should have mostly addressed the first concern. The process was done by matching the spectra visually and may be imperfect. The second source is the most likely contributor to differences between the two spectra. As outlined, the spectrograph was not thermally stabilized and the traces were observed to drift by up to several pixels over the course of a night. This made it difficult to flat field properly and also induced some errors in fitting a wavelength solution. If further attention was given to improving the thermal stability of the spectrograph and how to calibrate it (both in data extraction and reduction) we believe the noise could be reduced substantially. This is beyond the scope and the aims of this body of work. Finally, late type stars like mu Cepheus, and Alf Orionis are known to host shallow molecular absorption features, which vary in strength with time. This means it is difficult to compare spectra from these stars taken at different epochs and indeed some of the change in the spectra are due to the star itself. 

Spectra like these offer interesting insights into the composition of stars. Here we simply offer an overview of the types of features that can detected. The spectrum of HD182835, is typical for an F-type star, and is dominated by hydrogen absorption features. The hydrogen-Paschen transition is seen at $1282.16$~nm (black vertical line), by far the most prominent feature in J-band. Nine hydrogen-Brackett transitions across most of the H-band spectrum are also clearly visible (red vertical lines). mu Cepheus and Alpha Orionis are cooler late M-type stars, so the hydrogen features are not as prominent, and instead J and H-band spectra are dominated by the complicated blended molecular absorption features of CO lines (orange vertical lines), which are typically broader and shallower. The presence of magnesium (purple vertical lines) and silicon (brown vertical lines) can also be seen in the spectra of these stars. These features are well established and useful when measuring temperature and metallicities.

These results validate the basic operation of the photonic-based spectrograph and indeed the entire system. In addition, they demonstrate the practical implementation of the instrument for one possible application (i.e. stellar spectroscopy). These were the first spectra collected with a photonic spectrograph behind an AO system.  

\section{Conclusion}\label{sec:conclusion}
We have demonstrated a highly efficient photonic-based spectrograph that could be used for stellar or exoplanet spectroscopy (when combined with a science-grade detector). The instrument uniquely combines an extreme adaptive optics system and pupil apodization optics to achieve high coupling efficiency into a SMF, which can be efficiently dispersed across the J and H bands by using an AWG based spectrograph. The instrument currently offers a throughput of $5\%$ from sky-to-detector which we outline could easily be upgraded to $\sim13\%$ (at $1550$~nm assuming a coupling efficiency of $50\%$). The spectrograph had an internal efficiency of $42\%$, which can be improved to over $55\%$ simply by adding anti-reflection coatings to the bulk optics we used. Regardless, these levels of throughput are already competitive with current MMF fed spectrographs. The resolving power was $4000$--$5000$ across the J and H bands of operation and the entire instrument occupied a volume of less than $30\times30\times30$~cm$^{3}$. Such an instrument could be used to collect medium-resolution spectra of imagable exoplanets currently studied with integral field spectrographs~\cite{barman2015}. 

The possibility to break traditional instrument scaling laws and eliminate modal noise by using a SMF is a strong motivator to consider employing such solutions. In addition, commercially available AWG devices with similar specifications to the one presented here are relatively inexpensive (USD $\$1$--$2$k). This is the first time that an optimized and highly efficient photonic-based spectrograph has been realized and demonstrated for astronomical applications. This solution will no doubt play a key role in future instrument design, especially for instruments on ELTs.    

\section*{Funding}
Japan Society for the Promotion of Research ($23340051$, $26220704$, $23103002$). Australian Research Council Centre of Excellence for Ultrahigh bandwidth Devices for Optical Systems (CE$110001018$)


\section*{Acknowledgments} This work was supported by the Astrobiology Center (ABC) of the National Institutes of Natural Sciences, Japan and the directors contingency fund at Subaru Telescope. The authors wish to recognize and acknowledge the very significant cultural role and reverence that the summit of Maunakea has always had within the indigenous Hawaiian community. We are most fortunate to have the opportunity to conduct observations from this mountain.

\end{document}